\documentclass[10pt, twocolumn]{aastex701}

\let\tablenum\relax  
\usepackage{siunitx, amsmath, xfrac}
\usepackage{isomath, natbib}
\usepackage{mhchem, derivative}
\usepackage{needspace}

\usepackage[utf8]{inputenc}
\usepackage[T1]{fontenc}
\DeclareSIUnit{\Msun}{\mathnormal{M_\odot}}
\DeclareSIUnit{\Rsun}{\mathnormal{R_\odot}}
\DeclareSIUnit{\year}{yr}
\DeclareSIUnit{\erg}{erg}

\newcommand{\unsim}{\mathord{\sim}}
\newcommand{\Lag}[1]{\ensuremath{{\rm L}_{#1}}}

\begin{document}
	
\title[Contact-binary evolution with saturated magnetic braking]{Contact-binary evolution with energy transfer and saturated magnetic braking}

\author[0000-0003-4200-7852]{Matthias Fabry}
\affiliation{Department of Astrophysics and Planetary Science, 800 E. Lancaster Ave., Villanova, PA 19085, USA}
\email{matthias.fabry@villanova.edu}
\author[0000-0002-1913-0281]{Andrej Pr\v sa}
\affiliation{Department of Astrophysics and Planetary Science, 800 E. Lancaster Ave., Villanova, PA 19085, USA}
\email{andrej.prsa@villanova.edu}


\begin{abstract}
	The evolution of low-mass contact binaries is influenced by angular-momentum loss, mass and energy transfer, and the nuclear evolution of the components.
	They have periods shorter than one day, and we expect their period evolution to be dominated by magnetic braking.
	Evidence for saturated magnetic braking was presented by studying the period distribution of detached eclipsing binaries.
	This means the strength of magnetic braking likely does not cause a steep period-shrinking relation derived from the widely used Skumanich law.
	We find further evidence for saturated magnetic braking by considering evolutionary models of low-mass contact binaries.
	We also show that energy transfer must play an important role over a wide parameter range in producing the observed low mass ratios of contact binaries.
\end{abstract}

\section{Introduction}\label{sec:intro}
Low-mass contact binaries (also known as W UMa stars) have been studied since the first catalogs of variable stars were available \citep[e.g.,][]{eggenContactBinariesII1967, binnendijkOrbitalElementsUrsae1970}.
They are characterized by a smoothly variable light curve, indicating that the envelopes of both components overflow their respective Roche Lobes (RLs).
In the past decades, both ground- and space-based photometric monitoring surveys, such as OGLE \citep{szymanskiContactBinariesOGLEI2001}, \textsl{Hipparcos} \citep{selamKeyParametersUMatype2004}, ASAS-SN \citep{paczynskiEclipsingBinariesASAS2006}, \textsl{Kepler} \citep{boruckiKeplerPlanetDetectionMission2010}, and \textsl{TESS} \citep{rickerTransitingExoplanetSurvey2015} have detected tens of thousands of candidate contact binaries.

The evolution of W UMa stars is not fully understood.
Qualitatively, \citet{eggletonFormationEvolutionContact2012} describes that contact binaries form by orbital decay through interactions with a third body, and through tidal friction caused by magnetic braking (MB).
Once contact between the components is established however, no complete structure and evolution model is yet firmly established.
\citet{lucyStructureUrsaeMajoris1967} provided the first static model of contact binaries with energy transfer (ET), but it could not explain the period-color relation of W UMa stars.
In later works, the contact discontinuity model was put forward \citep{shuStructureContactBinaries1976, shuStructureContactBinaries1979}, but it was shown it violated the second law of thermodynamics \citep{papaloizouMaintenanceTemperatureDiscontinuity1979}.
\citet{flanneryCyclicThermalInstability1976} and \citet{lucyUrsaeMajorisSystems1976} developed a model that exhibits thermal relaxation oscillations (TROs).
They showed that the combination of energy and mass transfer in a contact binary causes a thermal cycle around a state of shallow contact \citep[see also][]{yakutEvolutionCloseBinary2005,liDynamicalStabilityUrsae2006}.
None of these models however present a complete evolutionary sequence, from first contact to the onset of the merger.

Yet, understanding the full evolution of W UMa binaries is important to unravel the conditions of their merger.
It is still uncertain whether the Darwin instability, occurring when the orbital angular momentum drops below one third of the spin angular momentum of both stars \citep{darwinDeterminationSecularEffects1879, hutStabilityTidalEquilibrium1980, rasioMinimumMassRatio1995} or unstable mass loss from the second Lagrangian point \citep{kuiperInterpretationLyraeOther1941, pejchaBuryingBinaryDynamical2014} is the starting point for the merger.
In the past two decades, several Luminous Red Novae (LRNe) have been identified and were associated with close binary mergers.
V1309 Sco erupted in 2008 \citep{masonPeculiarNovaV13092010}, and was shown by \citet{tylendaV1309ScorpiiMerger2011} to be a merger of a low-mass contact binary.
Since then, more LRNe were identified, but it is not yet clear whether a contact binary merger is the only progenitor channel, or if common-envelope ejections are possible too \citep{blagorodnovaCommonEnvelopeEjection2017, blagorodnovaProgenitorPrecursorEvolution2020, blagorodnovaLuminousRedNova2021}.

An important ingredient in the angular-momentum evolution of low-mass stars is magnetic braking \citep[MB, e.g.,][]{schatzmanTheoryRoleMagnetic1962, mestelMagneticBrakingStellar1968, mestelMagneticBrakingLatetype1987}.
Due to magnetic activity, the stellar wind is trapped and forced to co-rotate until a distance much larger than the stellar radius.
In single stars, MB causes the spin-down of the star over time, first recognized in G-type dwarfs by \citet{skumanichTimeScalesCa1972}.
The now-called Skumanich law relates the rotational velocity to stellar age by $v_{\rm rot} \sim t^{-1/2}$.
In close binaries ($p_{\rm orb} \lesssim \qty{5}{\day}$), we expect tides to synchronize the stellar rotation period to the orbital period \citep{zahnDynamicalTideClose1975}.
Therefore, if we assume efficient tidal synchronization in contact-binary progenitors, rather than the stars losing angular momentum, spin-orbit coupling causes a decrease in orbital angular momentum instead.
This, in turn, causes the period of the binary to shorten.

The strength of MB has important ramifications for the evolution of all close binaries, including Cataclysmic Variables \citep[CVs,][]{pattersonEvolutionCataclysmicLowmass1984, hameuryMagneticBrakingEvolution1988, warnerCataclysmicVariableStars2003, kniggeEvolutionCataclysmicVariables2011} and low-mass X-ray binaries \citep[LMXBs,][]{podsiadlowskiEvolutionarySequencesLow2002}.
In CVs and LMXBs, MB determines the mass-transfer rate and also their observable properties, such as the X-ray flux \citep[e.g.,][]{rappaportFormationEvolutionLuminous1994, ivanovaMagneticBrakingRevisited2003, vanLowmassXrayBinaries2019, dengEvolutionLMXBsDifferent2021}.

The exact mechanism of MB and its resulting angular-momentum loss (AML) is still poorly understood.
Different models produce different AML prescriptions.
The model of \citet{weberAngularMomentumSolar1967}, which agrees with the $t^{-1/2}$ dependence from \citet{skumanichTimeScalesCa1972}, and further developed by \citet{verbuntMagneticBrakingLowmass1981} and \citet{rappaportNewTechniqueCalculations1983}, causes a $\dot{J}_{\rm MB} \sim \Omega^3_{\rm rot}$ relation, and thus a $\dot{J}_{\rm MB} \sim p_{\rm orb}^{-3}$ relation in synchronized binaries.
On the other hand, studies show that, while there is a strong correlation between stellar activity and rotation rate, they also show this relation saturates at the highest of rotational velocities \citep[e.g.,][]{staufferDistributionRotationalVelocities1987, newtonHaEmissionNearby2017, johnstoneActiveLivesStars2021}.
\citet{keppensEvolutionRotationalVelocity1995} shows that a saturated dynamo model can explain the rotation-rate distribution of clusters such as the Pleiades and Hyades, in particular the rapid rotators of the Pleiades \citep{staufferDistributionRotationalVelocities1987}.
At high rotational velocities, the magnetic field strength ceases to increase with rotation rate, which allows for the survival of rapid rotators \citep[see also, e.g.,][]{quelozRotationalVelocityLowmass1998,andronovCataclysmicVariablesEmpirical2003}.
Saturated MB laws were used in studies by \citet{kawalerAngularMomentumLoss1988}, \citet{chaboyerStellarModelsMicroscopic1995}, \citet{sillsAngularMomentumEvolution2000} and \citet{mattMASSDEPENDENCEANGULARMOMENTUM2015} among others.

Recently, \citet{el-badryMagneticBrakingSaturates2022} provided strong evidence for saturated MB by studying the detached eclipsing binary sample from the Zwicky Transient Facility.
The observed period distribution could only be reconciled with a MB law that is linear in the period (rather than cubic as with classical MB).
In other recent work, disrupted MB has been considered as well.
In this model, MB is thought to shut off altogether if the star becomes fully convective.
Studying detached white-dwarf binaries with a K or M dwarf companion, \citet{belloniEvidenceSaturatedDisrupted2024} found that below masses of around $\qty{0.3}{\Msun}$, the observed post-common-envelope binaries can be explained if MB is (suddenly) very weak for these masses (although this need not be the case for evolved stars of such masses, see \citealp{belloniResolutionParadoxSDSS2025}).
Meanwhile, they required stronger MB than initial saturated prescriptions for stars with a radiative core.

In this work, we study the effect of MB and ET on the properties of low-mass contact binaries.
In particular, we investigate the duration of the contact phase under different MB laws, and assess the mass-ratio evolution under a model of ET.
This paper is structured as follows.
In Sect.~\ref{sec:analytic} we analytically compute the maximal lifetime of contact binaries under different MB laws.
Section \ref{sec:evol} presents the setup of the numerical models, while in Sect.~\ref{sec:res} we show and discuss the results from the computations.
Lastly, Sect.~\ref{sec:conc} gives some concluding remarks.

\section{Analytical considerations}\label{sec:analytic}
\begin{figure}
	\centering
	\includegraphics[width=\columnwidth]{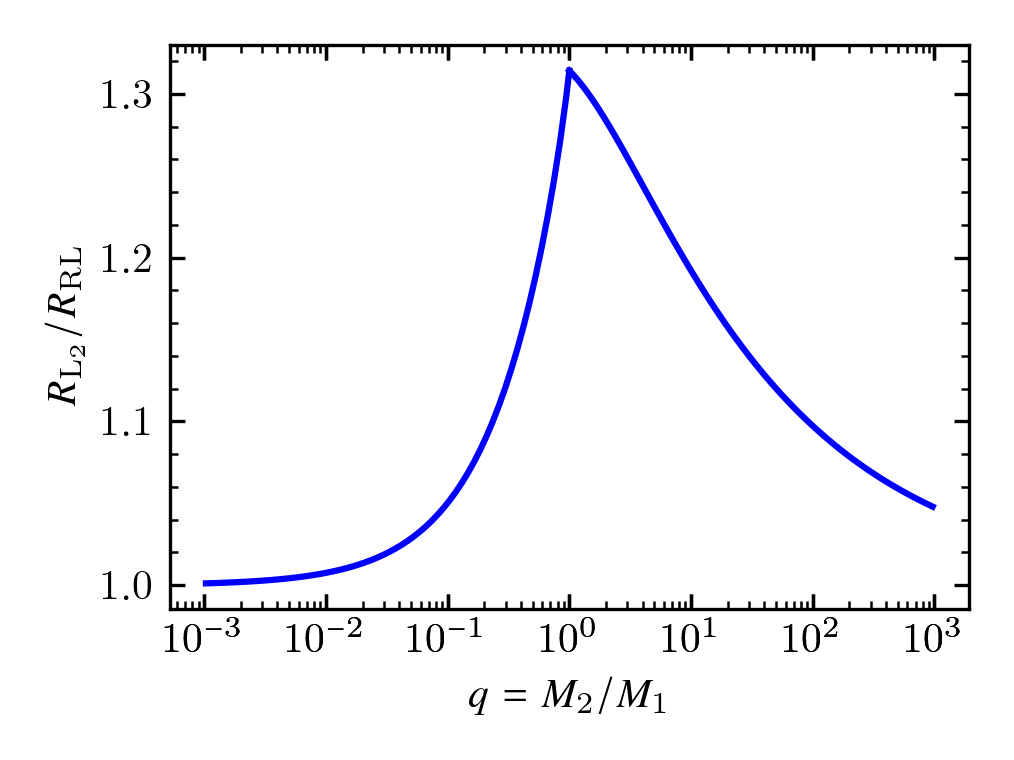}
	\caption{Volume-equivalent-radius ratio of the ${\rm L_2}$ and Roche Lobe equipotentials as a function of the mass ratio. This function is discontinuous at $q = 1$ because the location of ${\rm L_2}$ moves from the far side of $M_2$ to the near size of $M_1$.}
	\label{fig:l2ratios}
\end{figure}

As a low-mass binary reaches a contact configuration, we can make simple estimates of its lifetime under different prescriptions of AML.
We expect the binary to merge on a dynamical timescale once the common surface reaches the second Lagrangian point, \Lag{2} \citep[e.g.,][]{pejchaBuryingBinaryDynamical2014}.
From volume integrations of the Roche geometry \citep[][]{kopalCloseBinarySystems1959,fabryModelingOvercontactBinaries2022}, we have that the volume-equivalent radius of the \Lag{2} equipotential, $R_{\Lag{2}}$, is at most around 30\% larger than that of the Roche Lobe, $R_{\Lag{1}} \equiv R_{\rm RL}$, at the same binary separation, $a$, see Fig.~\ref{fig:l2ratios}.
The orbital angular momentum $J$ is
\begin{equation}
	J^2 = G\frac{M_1^2M_2^2}{M_1+M_2}a,
\end{equation}
for a circular orbit with masses $M_1$ and $M_2$, and $G$ being the gravitational constant.
Therefore, neglecting any mass transfer (MT) or radius changes, only around 12\% of the angular momentum needs to be dissipated for a binary that just entered in contact to reach the \Lag{2} surface.
This number would go down as the mass ratio, $q \equiv M_2/M_1$, departs from unity since $R_{\Lag{2}}(R_{\rm RL})$ is a decreasing function of $q$.
Thus, without MT, we expect the lifetime of a contact binary to be only around 10\% of the AML timescale $\tau_{\rm AML} = J / \dot{J}_{\rm AML}$, which can be estimated from the different descriptions of MB.
\par

For the non-saturated prescription by \citet[henceforth referred to as RVJ]{rappaportNewTechniqueCalculations1983}, applied to synchronized binaries, the MB torque is
\begin{equation}\label{eq:mb_rvj}
	\odv{J}{t}_{\rm RVJ} = -\qty{3.8e-30}{\s\squared\per\cm} MR_\odot^4\left(\frac{R}{R_\odot}\right)^{\gamma}\left(\frac{2\pi}{p_{\rm orb}}\right)^3.
\end{equation}
The parameter $\gamma$ can be adjusted to account for uncertainties, but is usually set to $\gamma = 4$, which we do here as well.
Scaling to solar values gives \citep{el-badryMagneticBrakingSaturates2022}:
\begin{equation}
	\odv{J}{t}_{\rm RVJ} = -\mathcal{\dot{J}}_0 \left(\frac{M}{M_\odot}\right)\left(\frac{R}{R_\odot}\right)^4\left(\frac{p_{\rm orb}}{\unit{\day}}\right)^{-3},
\end{equation}
where $\mathcal{\dot{J}}_0 = \qty{6.8e34}{\erg}$ is the calibration constant.
In this work, we also refer to this prescription of MB as ``classical MB.''
Therefore, a system consisting of two $\qty{1}{\Msun}$, $\qty{1}{\Rsun}$ stars in a $\qty{0.5}{\day}$ orbit (equivalent to a separation of $\qty{3.34}{\Rsun}$) has an expected lifetime of
\begin{equation}\label{eq:tau_rvj}
	\tau_{\rm RVJ} \approx 0.1 \frac{\sqrt{\frac{GM_\odot^3a}{2}}}{2\mathcal{\dot{J}}_0(0.5)^{-3}} = \qty{2.3e7}{\year}.
\end{equation}
This number is an upper limit of the contact-binary lifetime because $\dot{J}$ increases with decreasing period.
Note that $\tau_{\rm RVJ}$ is of the same order as the thermal timescale of solar-type stars.
\par

Saturated MB was parameterized by \citet{sillsAngularMomentumEvolution2000} as
\begin{equation}\label{eq:mb_sat}
	\odv{J}{t}_{\rm sat} = \begin{cases}
		-\mathcal{\dot{J}}_1\!\left(\frac{M_\odot}{M}\frac{R}{ R_\odot}\right)^{\frac{1}{2}}\!\left(\frac{p_{\rm orb}}{\unit{\day}}\right)^{-3},\\
		\hfill p_{\rm orb} > p_{\rm crit} \\
		-\mathcal{\dot{J}}_1\!\left(\frac{M_\odot}{M}\frac{R}{R_\odot}\right)^{\frac{1}{2}}\!\left(\frac{p_{\rm orb}}{\unit{\day}}\right)^{-1}\!\left(\frac{p_{\rm crit}}{\unit{\day}}\right)^{-2},\\
		 \hfill p_{\rm orb} \leqslant p_{\rm crit}
\end{cases}
\end{equation}
where $\mathcal{\dot{J}}_1 = \qty{1.04e35}{\erg}$ is another calibration constant, and $p_{\rm crit}$ is the mass-dependent critical period where saturation occurs.
We follow \citet{el-badryMagneticBrakingSaturates2022}, who adopted the empirical fit of \citet{wrightSTELLARACTIVITYROTATIONRELATIONSHIP2011}:
\begin{subequations}
	\begin{align}
		p_{\rm crit} &= 0.1p_{\rm rot, \odot}\tau_{\rm conv},\\
		\log \tau_{\rm conv} &= -1.49\log \frac{M}{M_\odot} - 0.54 \log^2\frac{M}{M_\odot}.
	\end{align}
\end{subequations}

With $p_{\rm rot,\odot} = \qty{28}{\day}$, the critical period of a $\qty{1}{\Msun}$ star is $\unsim\qty{2.8}{\day}$.
Therefore, for the same $\qty{1}{\Msun}$, $\qty{1}{\Rsun}$ twin binary with $p_{\rm orb} = \qty{0.5}{\day}$, the MB strength is about one order of magnitude weaker, which results in
\begin{equation}\label{eq:tau_s00}
	\tau_{\rm S00} \approx 0.1 \frac{\sqrt{\frac{GM_\odot^3a}{2}}}{2\mathcal{\dot{J}}_1(0.5)^{-1}(2.8)^{-2}} = \qty{4.6e8}{\year}.
\end{equation}
This is longer than the typical thermal timescale of solar-type stars, but still shorter than their nuclear one.
\par

Given that we expect contact binaries to evolve on their nuclear timescale, several processes must counteract the AML to MB.
MT from a low- to a high-mass component can achieve this through two effects.
First, from the conservation of total angular momentum, the period increases as mass is transferred from a lower to a higher mass component.
The second effect is through spin-orbit coupling.
Assuming good tidal synchronization, it injects angular momentum back into the orbit since matter flows to locations closer to the barycenter.
In Sect.~\ref{sec:res}, we show that only strong ET can drive MT from the lower mass to the higher mass component in W UMa stars.
\par

\section{Numerical evolution models}\label{sec:evol}
To investigate the behavior of close binaries under different description of MB and ET, we compute binary-evolution models using \texttt{MESA} \citep{paxtonModulesExperimentsStellar2011, paxtonModulesExperimentsStellar2013, paxtonModulesExperimentsStellar2015, paxtonModulesExperimentsStellar2018, paxtonModulesExperimentsStellar2019, jermynModulesExperimentsStellar2023}, version 24.03.1.
We focus on the main-sequence evolution of solar-type stars in close binary systems that evolve to become contact binaries.
We simulate systems with initial primary mass in the range $M_{1, \rm init} = \qtyrange{0.5}{1.4}{\Msun}$, with a $\qty{0.1}{\Msun}$ spacing, initial period between $p_{\rm init} = \qtyrange{0.3}{5}{\day}$, logarithmically spaced with $\Delta \log p_{\rm init}/\unit{\day} = \num{0.07}$, and initial mass ratios between $q_{\rm init} = \numrange{0.35}{0.95}$, spaced with $\Delta q_{\rm init} = 0.1$.
All models are computed at solar metallicity, $Z_{\odot} = 0.0142$ \citep{asplundChemicalCompositionSun2009}, with metal fractions also following \citet{asplundChemicalCompositionSun2009}.
Below we outline the used microphysics, binary interaction physics and assumption on angular momentum evolution.

\subsection{Microphysics and mixing}
We use the basic nuclear network with the isotopes $\ce{^1H, ^3He, ^4He, ^{12}C, ^{14}N, ^{16}O, ^{20}Ne}$ and $\ce{^{24}Mg}$, appropriate to follow the main-sequence evolution.
Reaction rates are taken from the JINA REACLIB \citep{cyburtJINAREACLIBDatabase2010} and NACRE \citep{anguloCompilationChargedparticleInduced1999} databases, with weak reactions added from \citet{fullerStellarWeakInteraction1985}, \citet{odaRateTablesWeak1994} and \citet{langankeShellmodelCalculationsStellar2000}.
A prescription of nuclear screening is included following \citet{chugunovCoulombTunnelingFusion2007}, and thermal-neutrino losses are computed from \citet{itohNeutrinoEnergyLoss1996}.
\par

The \texttt{MESA} equation of state is computed from a blend of tables from the OPAL \citep{rogersUpdatedExpandedOPAL2002}, SCVH \citep{saumonEquationStateLowMass1995}, HELM \citep{timmesAccuracyConsistencySpeed2000}, PC \citep{potekhinThermodynamicFunctionsDense2010} and Skye \citep{jermynSkyeDifferentiableEquation2021} projects.
The blend prescription is detailed in \citet{jermynModulesExperimentsStellar2023}.
\par

Radiative opacities are taken from OPAL \citep{iglesiasRadiativeOpacitiesCarbon1993, iglesiasUpdatedOpalOpacities1996}, with low-temperature data from \citet{fergusonLowTemperatureOpacities2005} and high-temperature, Compton-dominated regime by \citet{poutanenRosselandFluxMean2017}.
Electron conductivity is included following \citet{cassisiUpdatedElectronconductionOpacities2007} and \citet{blouinNewConductiveOpacities2020}.
\par

Convective regions are determined from the Ledoux criterion \citep{ledouxStellarModelsConvection1947}.
Convective mixing in such regions is modeled following mixing length theory \citep{vitenseWasserstoffkonvektionszoneSonneMit1953, bohm-vitenseUberWasserstoffkonvektionszoneSternen1958}, using the prescription of \citet{coxPrinciplesStellarStructure1968}, with a mixing length parameter of $\alpha = 1.5$.
A small exponential overshoot region is added following \citet{herwigEvolutionAGBStars2000}, where the diffusion coefficient $0.001$ pressure-scale heights within the convective region decays exponentially outward with an e-folding length of $0.025$ pressure-scale heights.
Thermohaline mixing is included following \citet{kippenhahnTimeScaleThermohaline1980}, with an efficiency of unity.
We include rotational mixing induced by Eddington-Sweet circulation \citep{eddingtonCirculatingCurrentsRotating1925, sweetImportanceRotationStellar1950}, the Goldreich-Schubert-Fricke instability \citep{goldreichDifferentialRotationStars1967, frickeInstabilitatStationarerRotation1968}, and the secular- and dynamical shear instabilities following \citet{zahnCirculationTurbulenceRotating1992} and \citet{endalEvolutionRotatingStars1978}.
\par

Although solar-type main-sequence stars have a weak wind, with a negligible effect on their evolution, we include the simple wind mass-loss prescription from \citet{reimersCircumstellarEnvelopesMass1975}.
\par

\subsection{Angular momentum and tides}\label{ssec:ang_mom_tides}
We model AML from gravitational radiation (GR) from the quadrupole component \citep[e.g.,][]{landauClassicalTheoryFields1975}:
\begin{align}
	-\frac{1}{J}\odv{J}{t}_{\rm GR} &= \frac{32G^3}{5c^5}\frac{M_1M_2(M_1+M_2)}{a^4} \\
		&= \frac{ \mathcal{\dot{J}}_{\rm GR}}{J}\left(\frac{M_1}{\unit{\Msun}}\right)^{\frac{10}{3}}\left(\frac{p_{\rm orb}}{\unit{\day}}\right)^{\frac{-7}{3}}\frac{q^2}{\left(1+q\right)^{\frac{2}{3}}}
\end{align}
with $c$ the speed of light and $\mathcal{\dot{J}}_{\rm GR} \approx \qty{1.04e33}{\erg}$ the AML scale of GR for solar-type contact binaries.
\par

The effect of AML from MB is modeled following the prescriptions of \citet{rappaportNewTechniqueCalculations1983} and \citet{sillsAngularMomentumEvolution2000}, quantified in Eqs.~\eqref{eq:mb_rvj} and \eqref{eq:mb_sat}, respectively.
MB requires a sufficiently extended convective envelope to cause magnetic activity \citep[e.g.,][]{pylyserEvolutionLowmassClose1988}.
Following \citet{podsiadlowskiEvolutionarySequencesLow2002}, for convective envelopes with mass fractions smaller than $\chi = m/M < 0.02$, we scale both MB prescriptions by a factor
\begin{equation}\label{eq:mb_factor}
	f_{\rm conv} = \exp\left(1-0.02/\chi_{\rm conv}\right).
\end{equation}
Additionally, we shut off MB for a component if either its convective envelope contains more than 99\% of the stellar mass, or if a convective core with a mass fraction over 5\% develops.
\par

We expect close binaries to be efficiently synchronized to their Keplerian velocity \citep{zahnDynamicalTideClose1975, zahnTidalFrictionClose1977}.
Hence we operate a tidal synchronization prescription on the timescale of the orbit to uniformly synchronize the rotation rate of the components to the orbital period:
\begin{equation}
	\Delta{j}_{\rm tides} = \left[1-\exp\left(\frac{-\Delta{t}}{p}\right)\right](\omega_{\rm orb}i_{\rm rot} - j).
\end{equation}
Here $i_{\rm rot}$ is the moment of inertia of the stellar layer, $j$ its specific angular momentum, and $\omega_{\rm orb} = 2\pi/p$ is the orbital angular velocity.
In practice, for our short-period binaries, this means there is little departure of the rotation rate from the Keplerian velocity, with almost no differential rotation.
Tides couple the orbital angular momentum of the orbit and the spin angular momentum of the components, which we account for by adding (subtracting) the amount of angular momentum that was removed (added) by the above tidal prescription.
\par

\subsection{Binary interaction}\label{ssec:binary_methods}
Given that the components in contact binaries are highly distorted by tides, we model this influence by solving the stellar-structure equations in coordinates of constant Roche potential, $\Psi$, following the methodology of \citet{kippenhahnSimpleMethodSolution1970} and \citet{endalEvolutionRotatingStars1976}.
The independent coordinate is changed to a volume-equivalent radius, and adjusting the equations with the appropriate correction factors, which are computed in \citet{fabryModelingOvercontactBinaries2022}.
With this approach, we can consistently model contact phases of arbitrary overflow, and together with energy transfer detailed below, we ensure shellularity of the common layers, meaning no baroclinicity occurs there.
\par

We assume fully conservative MT in our evolutionary models, with the MT rate computed implicitly so that, in a semi-detached configuration, the donor star just fills its RL, or, in a contact configuration, that the surfaces of the two components satisfy:
\begin{equation}
	r_i = F(r_i, q),
\end{equation}
where we define
\begin{equation}
	r_i \equiv \frac{R_i - R_{i, \rm RL}}{R_{i, \rm RL}}
\end{equation}
the fractional RL overflow, and $q\equiv M_2/M_1$ the mass ratio.
The function $F$ is constructed directly from Roche potential integrations of \citet{fabryModelingOvercontactBinaries2022}.
\par

Finally, ET is modeled using the method of \citet{fabryModelingContactBinaries2023}, by adjusting the luminosity of the stellar layers in the vicinity of the RL so that each component has the same brightness.
Because the convective envelopes of low-mass stars have a weaker dependency on gravity darkening ($T\sim g^{0.08}$, \citealt{lucyGravityDarkeningStarsConvective1967}, as opposed to $T\sim g^{1/4}$ following \citealt{vonzeipelRadiativeEquilibriumRotating1924}, for radiative envelopes), the luminosities just above the ET region, $L_i'$, are computed as
\begin{subequations}
	\begin{align}
		L_1'&= (L_1 + L_2)\frac{S_1}{S_1 + S_2},\\
		L_2'&= (L_1 + L_2)\frac{S_2}{S_1 + S_2}.
	\end{align}
\end{subequations}
Contrary to \citet{fabryModelingContactBinaries2023} and \citet{fabryModelingContactBinaries2025}, we do not apply numerical time-step smoothing and instead opt to take smaller evolutionary time-steps in the models to ensure numerical stability.
This is done to avoid introducing departures from shellularity during the integration, especially when the models are thermally oscillating, which was no issue for the radiative envelopes of the massive-star models in \citet{fabryModelingContactBinaries2023}.
\par

The width of the ET region in a contact binary is highly uncertain.
\citet{shuStructureContactBinaries1979} argue from order of magnitude estimates that the width, $d$, of the energy and mass interchange layer is small (of order $\delta \propto d/a \sim 10^{-2}$), which justified their use of a contact discontinuity.
This estimate however is shown to become quite large, or even exceed unity, for (very) massive stars \citep{fabryModelingContactBinaries2023}.
In W UMa stars, light curve analyses indicate that ET occurs also at the surface of the contact binary, rather than being a fully internal process \citep{zhouExplanationLightCurveAsymmetries1990}.
Some theoretical models also incorporate circulating surface flows \citep{kahlerStructureContactBinaries2004,stepienLargescaleCirculationsEnergy2009}.
However, since we cannot directly model surface flows in a 1D representation of the components, we base ourselves here on the properties of the thermally oscillating models.
Previous computations show that they oscillate around a state of shallow contact, with overflows typically less than $r < 0.05$ \citep{flanneryCyclicThermalInstability1976, liStructureEvolutionLowmass2004, yakutEvolutionCloseBinary2005}.
We therefore elect to perform the ET in the layers just above and below the RL, where the fractional overflow satisfies
\begin{equation}\label{eq:et_width}
	|r| \leqslant 0.02.
\end{equation}
We also smoothly ``turn on'' ET when $0 < \min(r_1, r_2) < 0.02$, by multiplying the to-be-transferred energy with a factor
\begin{equation}
	f = \frac{\min(r_1, r_2)}{0.02}.
\end{equation}
\par

\subsection{Termination}\label{ssec:termination}
The numerical models are halted according to different termination conditions:
\begin{itemize}
	\item No case A interaction. The models do not exchange mass while both stars are on the main sequence. For the period range examined here, these models follow case B evolution, in which we do not expect long-lived contact binaries.
	\item \Lag{2} overflow. The surface of the models reaches the \Lag{2} equipotential. When this happens, we expect (non-conservative) outflows from the outer Lagrangian point, which is associated with high AML, and we expect a merger within a dynamical timescale \citep{pejchaBuryingBinaryDynamical2014}.
	\item High MT rate. The implicit scheme solved for a MT rate above $\dot{M} = \qty{e-3}{\Msun\per\year}$. This is well above the thermal-timescale MT rate for solar-type stars, and we expect the system to transition to dynamical-timescale evolution. As \texttt{MESA} is not setup to compute this type of evolution, we halt the integration.
	\item Survival. The models undergo case A MT, and one star reaches the end of the main sequence before merging. This commonly happens for high-mass stars \citep{menonDetailedEvolutionaryModels2021,hennecoContactTracingBinary2024,fabryModelingContactBinaries2025}.
	\item TROs. The system underwent at least twenty TROs. Ideally one would like to follow the evolution until another termination condition is met, but this computationally expensive for a whole grid of models as the thermal timescale of the stars need to be resolved for a significant part of the main-sequence lifetime of the system. Therefore, in this work, we follow the system through a limited amount of TROs only, as it is sufficient to study the effects of the different ET and AML prescriptions during the contact phases of the models.
\end{itemize}

\section{Results and discussion}\label{sec:res}
The data required to reproduce the results in this Sect.~are available on \dataset[Zenodo]{\doi{10.5281/zenodo.16854984}}.
\subsection{Example models}\label{ssec:example}
\begin{figure}
	\centering
	\includegraphics[width=\columnwidth]{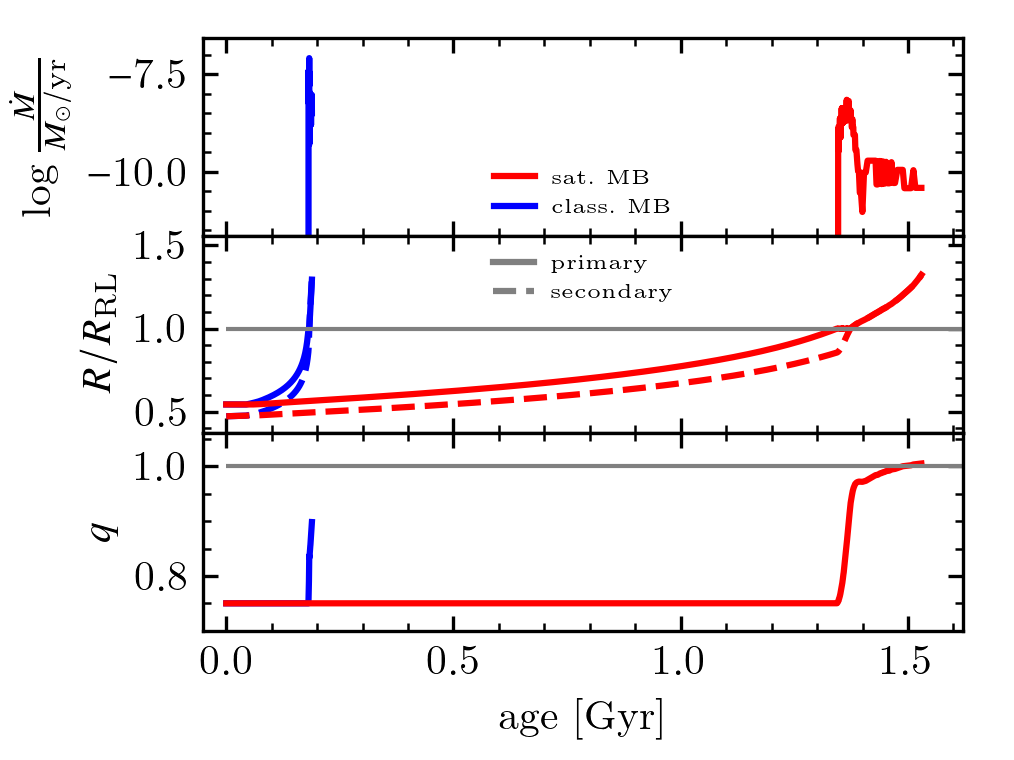}
	\caption{Mass-transfer rate, radius, and mass-ratio evolution of the $M_{\rm 1, init}=\qty{0.9}{\Msun}$, $p_{\rm init} = \qty{0.67}{\day}$, $q_{\rm init} = 0.75$ models without ET under the different MB laws. We note that the saturated MB model has its total lifetime as well as its time in binary interaction increased by about an order of magnitude.}
	\label{fig:mb_example}
\end{figure}

To visualize the effects of ET and the different MB prescriptions, we start this Sect.~by showing some representative example models.
In Fig.~\ref{fig:mb_example}, we show the evolution of the $M_{\rm 1, init}=\qty{0.9}{\Msun}$, $p_{\rm init} = \qty{0.67}{\day}$, $q_{\rm init} = 0.75$ models under both the Skumanich, classical and saturated MB laws.
No ET has been applied in these models.
While both these models are expected to merge on the main sequence, the timescales on which this happens is very different.
With classical MB, the model starts MT after around $\qty{0.2}{\giga\year}$, and merges promptly after, being unable to equalize the mass ratio.
On the other hand, with saturated MB, the model more gradually approaches the RL over about $\qty{1.35}{\giga\year}$, and has a binary interaction phase of around $\qty{0.2}{\giga\year}$ before it merges as a near-equal mass ratio contact binary.
From simple timescale arguments, the saturated MB model would be around an order of magnitude more likely to observe as the classical MB one, as expected from our estimations in Sect.~\ref{sec:analytic}.

\begin{figure}
	\centering
	\includegraphics[width=\columnwidth]{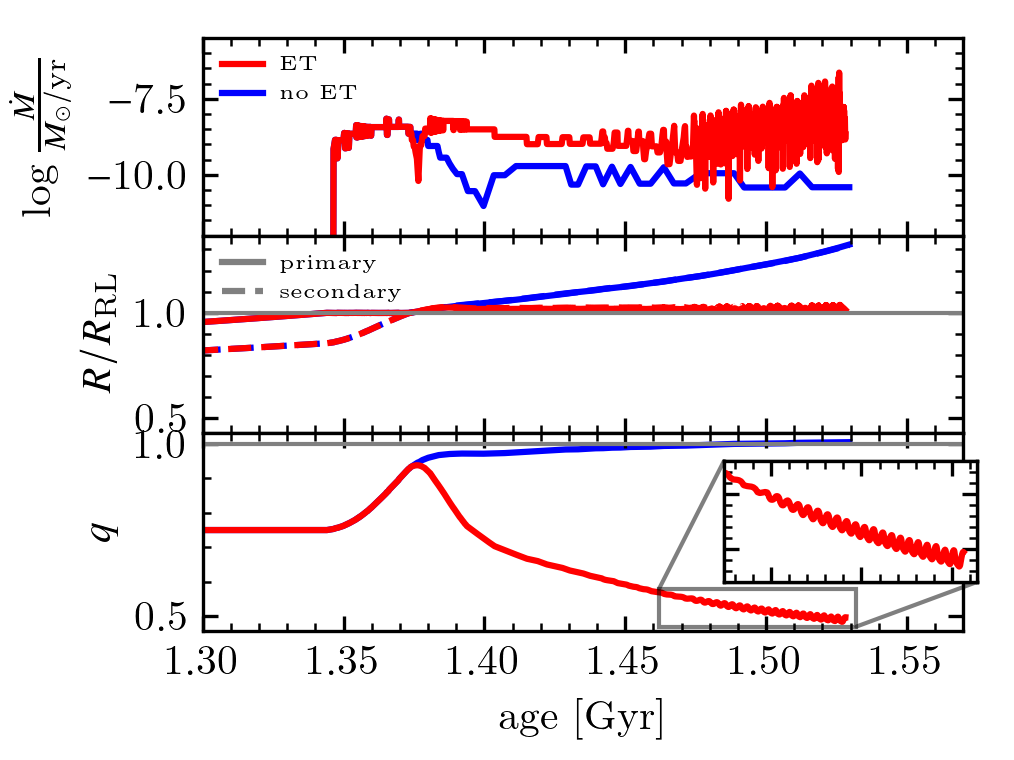}
	\caption{Mass-transfer rate, radius and mass-ratio evolution of the $M_{\rm 1, init}=\qty{0.9}{\Msun}$, $p_{\rm init} = \qty{0.67}{\day}$, $q_{\rm init} = 0.75$, with saturated MB, with ET (red curves) and without ET (blue curves). Without ET, the system merges at mass ratio of unity after MT is engaged, while the model with ET evolves away from mass ratio unity, and develops TROs (see inset).}
	\label{fig:example}
\end{figure}

In Fig.~\ref{fig:example}, we show the late-time evolution of the models with the same initial parameters as in Fig.~\ref{fig:mb_example}, only here both underwent saturated MB, and we comparing the ET model (red) with the no-ET model (blue).
Both models start MT at around $\qty{1.35}{\giga\year}$, and both reach mass ratios of around unity during this thermal-timescale MT phase.
Once contact is engaged, the no-ET model significantly reduces its MT rate, and evolves on the nuclear timescale toward \Lag{2} overflow.
The ET model evolves very differently, because the size of envelope of the energy gainer is very sensitive to the ET rate.
MT reverses and drives the mass ratio away from unity, in this case to values near 0.5 within the same time that the no-ET model merged.
During this phase, small oscillations develop, which become larger in amplitude toward the end of the simulation.
It is expected that those would continue to grow and cause the system to detach.
Whether the system would reach the Darwin instability first, or have its TROs become so large that \Lag{2} overflow would occur first is difficult to assess from these models, and it will depend very much on the assumed ET prescription (see below).

We note that, unlike \citet{stepienLargescaleCirculationsEnergy2009} and \citet{stepienEvolutionLowMass2012}, we do not require the reversal of mass ratio for contact binaries to exist.
They argue that, after the initial thermal disequilibrium of the first MT event, the secondary would again shrink within its RL and detach, akin to massive Algol systems.
However, in this example we see that, due to continuous ET, the secondary envelope is never allowed to thermally relax, and it becomes the donor in the system.
Only in larger-period systems is the initial MT phase long enough that reversal of mass ratio occurs.

\subsection{Grid outcomes}
\begin{figure*}
	\centering
	\includegraphics[width=.9\textwidth]{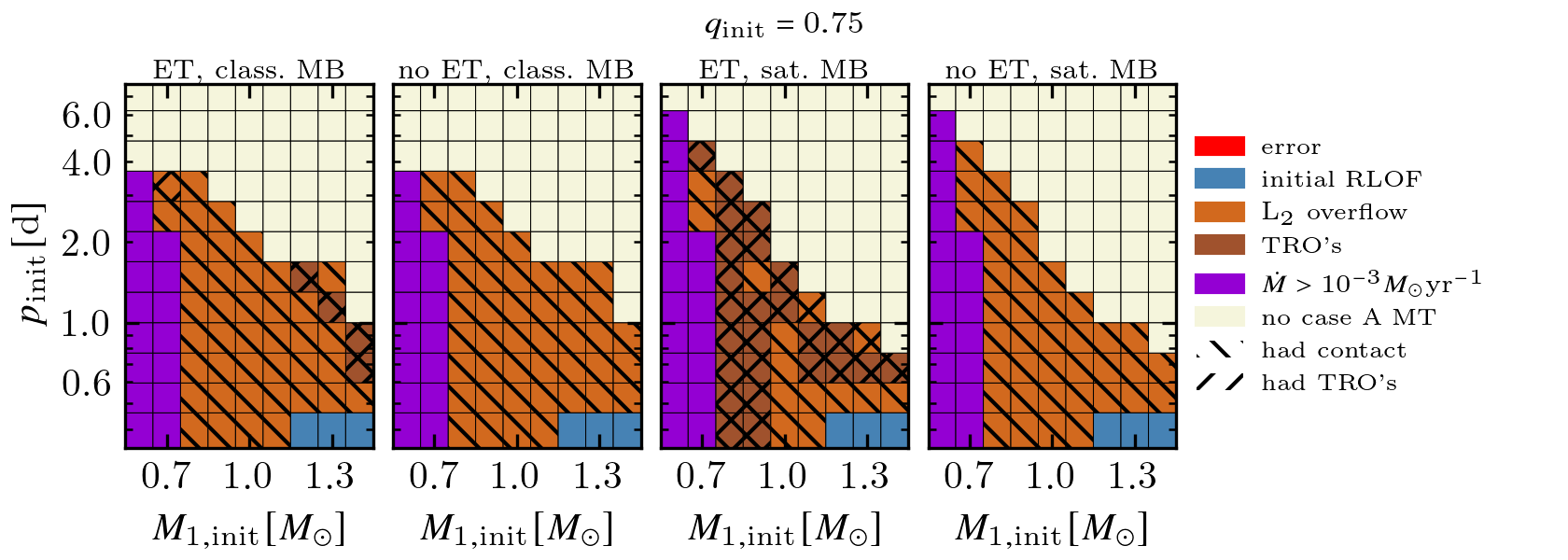}
	\caption{Termination of the $q=0.75$ models of our grids as function of initial primary mass $M_{1, \rm init}$ and initial period $p$. Termination conditions are listed in Sect.~\ref{ssec:termination}. We note that only the combination of ET and saturated MB allows for models over a large parameter range that experience TROs.}
	\label{fig:termination}
\end{figure*}

In Fig.~\ref{fig:termination}, we show the termination conditions of the $q = 0.75$ slice of the grid.
We distinguish four clusters of models.
First, at large period and/or mass, the models avoid case A MT, because MB is too weak to reduce the RL size enough before the primary star reaches the terminal-age main sequence.
At $\qty{1.4}{\Msun}$, there is hardly a convective envelope, rendering MB ineffective and making GR the dominant AML mechanism.
Here we note also that, contrary to massive stars modeled in \citet{menonDetailedEvolutionaryModels2021} or \citet{fabryModelingContactBinaries2025}, there are no main-sequence survivors due to strong tidal friction from MB, even in the saturated prescription.
Therefore, all systems that exchange mass in the explored parameter space merge during the main sequence, either due to runaway mass loss, or by eventually reaching the Darwin instability.

Second, at very low mass, the models suffer unstable MT, as the MT rate climbs above $\dot{M} = \qty{e-3}{\Msun\per\year}$.
This is due to the thick convective envelopes of these stars responding adiabatically to MT.
\citet{geAdiabaticMassLoss2015} find that for an $\qty{0.71}{\Msun}$ main-sequence star, the critical mass ratio\footnote{Note that \citet{geAdiabaticMassLoss2015} defined $q=M_{\rm donor}/M_{\rm accretor}$. We inverted their computed critical mass ratios to be consistent with our $q = M_{\rm accretor}/M_{\rm donor} = M_2/M_1$ definition.} below which dynamical MT would start is $q_{\rm crit} < 0.791$, while for an $\qty{0.8}{\Msun}$ main-sequence star they calculated $q_{\rm crit} < 0.696$.
Seeing that at $q_{\rm init} = 0.75$, the $\qty{0.8}{\Msun}$ model is stable, while the $\qty{0.7}{\Msun}$ model is not, we thus recover the approximate location of the stability boundary for adiabatic expansion under the assumption of conservative MT.

Third, the bulk of the models that either use the classical MB law or those that do not include ET merge during the main sequence, by reaching \Lag{2} overflow.
AML or nuclear expansion of the donor starts thermal timescale case A MT, leading to a contact binary with a mass ratio close to unity, which expands to the \Lag{2} equipotential.

Finally, in the model grid with ET and saturated MB, instead of merging due to \Lag{2} overflow, a large number of models hit the TRO termination condition.

\needspace{5\baselineskip}
\subsection{Expected lifetimes}
\begin{figure*}
	\centering
	\includegraphics[width=.9\textwidth]{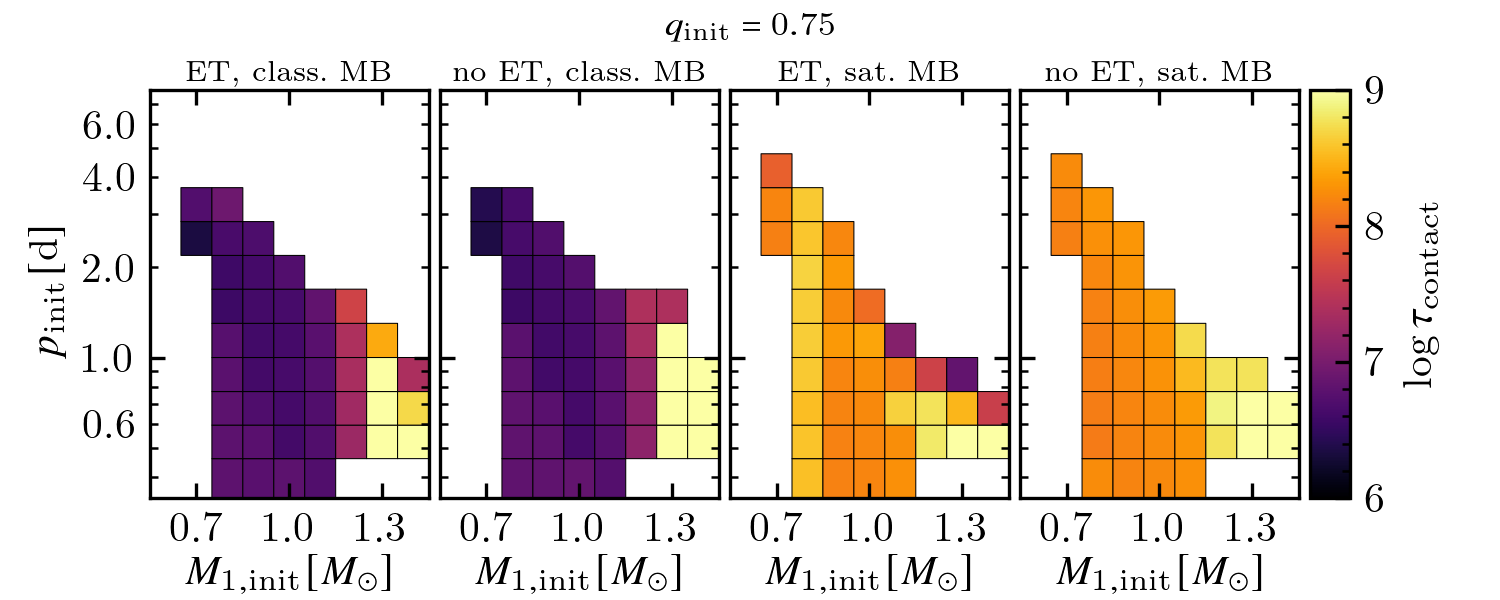}
	\caption{Lifetimes of the contact phases of the models with $q_{\rm init} = 0.75$. We confirm our estimates from Sect.~\ref{sec:analytic} that models using the classical MB law have an lifetime of only $\qty{e7}{\year}$, while saturated MB allows for lifetimes larger than $\qty{e8}{\year}$. The fact that the ET and no-ET models have largely the same lifetime is coincidental. For the models undergoing TROs (see Fig.~\ref{fig:termination}), the lifetimes reported here are lower limits.}
	\label{fig:taus}
\end{figure*}

In Sect.~\ref{sec:analytic} we estimated analytically that, when contact binaries suffer AML from MB according to the classical, Skumanich-type law, they would live at most for around a thermal timescale.
In Fig.~\ref{fig:taus}, we show the time spent in contact of the same $q_{\rm init}=0.75$ slice of our grid as in Fig.~\ref{fig:termination}.
We see that the models with classical MB only spend around $\qty{e7}{\year}$ in contact, as expected from our estimate in Eq.~\eqref{eq:tau_rvj}.
The exception to this are the models with $M_{\rm init} > \qty{1.2}{\Msun}$, where MB rapidly becomes inefficient due to their small convective envelopes.
ET has almost no effect on the $M < \qty{1.2}{\Msun}$ models, as the rate of AML is simply too high for a thermal response of the envelope to develop.

\begin{figure}
	\centering
	\includegraphics[width=\columnwidth]{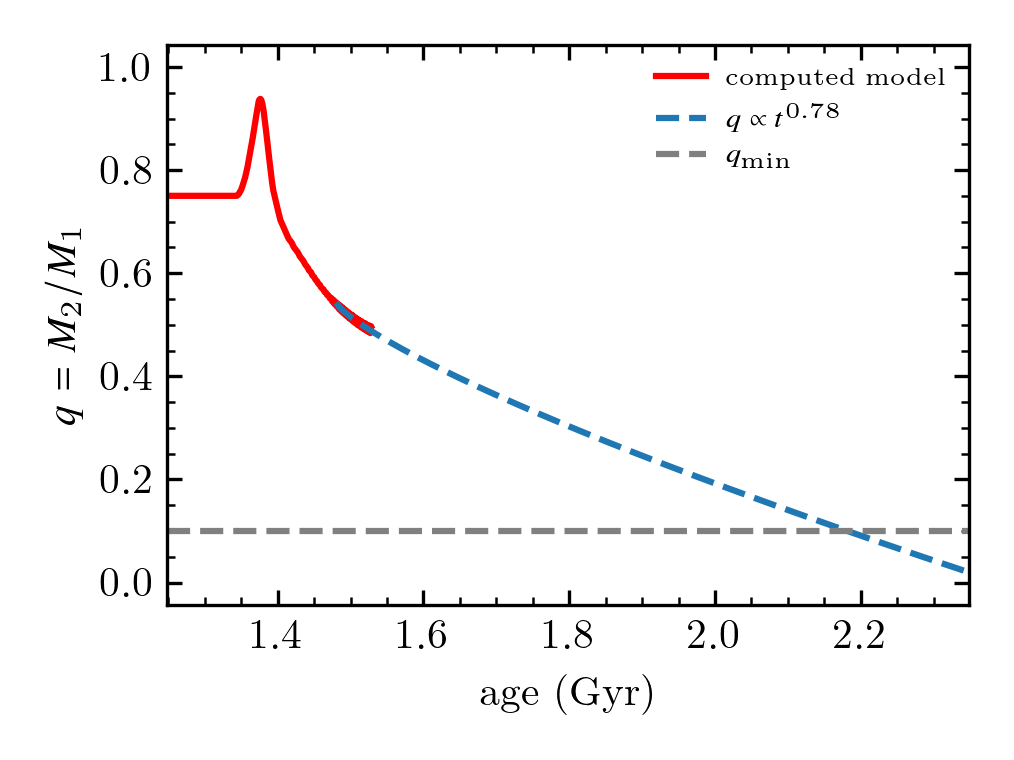}
	\caption{Mass-ratio extrapolation of the example model of Sect.~\ref{ssec:example}. We find that, if the system remains stable, it would reach the critical mass ratio $q_{\rm min}$ after around $\qty{0.7}{\giga\year}$ in contact.}
	\label{fig:extrapolation}
\end{figure}

From the right-most panel of Fig.~\ref{fig:taus}, we see that the model featuring saturated MB without ET have a lifetime of around $\qty{e8}{\year}$, aligning well with our estimate in Eq.~\eqref{eq:tau_s00}.
However, in the ET models that experience TROs, the reported numbers are lower limits as the simulations were cut short at an (arbitrary) 20 TROs.
It is expected that these systems will live longer, until either the Darwin instability is hit, or until the amplitude of the TROs brings the surface of the contact binary to the \Lag{2} equipotential.
A crude power-law extrapolation (Fig.~\ref{fig:extrapolation}) shows the mass ratio of the example model of Sect.~\ref{ssec:example} will reach the critical value of around $q_{\rm min} \approx 0.1$ \citep{rasioMinimumMassRatio1995} at around $\qty{2.1}{\giga\year}$.
The total time in contact will thus be around $\qty{0.7}{\giga\year}$, but this is highly dependent on how the radius-evolution proceeds, and if the MT rate remains stable.

\subsection{Mass-ratio distribution}
To predict the mass-ratio distribution from our model grids, we employ the same population-synthesis methods as in \citet{fabryModelingContactBinaries2025}.
This amount to computing the probability density
\begin{equation}
	\mathcal{P}(p', q') \propto \int \delta(q - q')\delta(p - p')\mathcal{W}_\mathcal{N}\odif{\mathcal{N}},
\end{equation}
where we sum the contributions of all models $\mathcal{N}$ when they are observed at period $p'$ and mass ratio $q'$.
We only change the prior distributions of initial primary mass, initial mass ratio and initial period, $\mathcal{W}_\mathcal{N} = \mathcal{W}_{M_1} \mathcal{W}_p \mathcal{W}_q$.
Following \citet{chabrierGalacticStellarSubstellar2003}, we take a Salpeter initial mass function for $M > \qty{1}{\Msun}$, and log-normal for $M < \qty{1}{\Msun}$.
The period distribution is log-normal and peaks at $\log(p_{\rm init}) \approx 5$ with a spread of $2.28$ \citep{raghavanSURVEYStelLARFAMILIES2010}.
According to \citet{moeMindYourPs2017}, the initial mass ratio of solar-type binaries has a negatively sloped power-law shape with a significant excess twin fraction at periods shorter than $\qty{10}{\day}$.
Hence we take as weighting functions for our models:
\begin{subequations}
	\small
	\begin{align}
		W_{M_1} dM_1 &=
		\begin{cases}
			M_1^{-2.3}\odif{M_1} & M_1 > \qty{1}{\Msun},\\
			A \exp\left(-\frac{(\log M_1 +0.65)^2}{2(0.57^2)}\right)\odif{\log M_1} & M_1 < \qty{1}{\Msun}.
		\end{cases}\\
		W_p \odif{\log p} &= \exp\left(-\frac{\log p - 5.03}{2.28^2}\right)\odif{\log p},\\
		W_q dq &= 
			\begin{cases}
				a q^{-0.5}     & q < 0.95, \\
				a q^{-0.5} + 6 & q\geqslant 0.95,
			\end{cases}
	\end{align}
\end{subequations}
Here $A = \num{4.4e-2}$ is a constant ensuring the piecewise functions of the mass distribution are continuous, and $a = 1.4/(1 - \sqrt{0.3})$ so that the twin fraction of $\mathcal{F_{\rm twin}} = 0.3$ is respected \citep[][their Eq.~6]{moeMindYourPs2017}.

\begin{figure*}
	\centering
	\includegraphics[width=.9\textwidth]{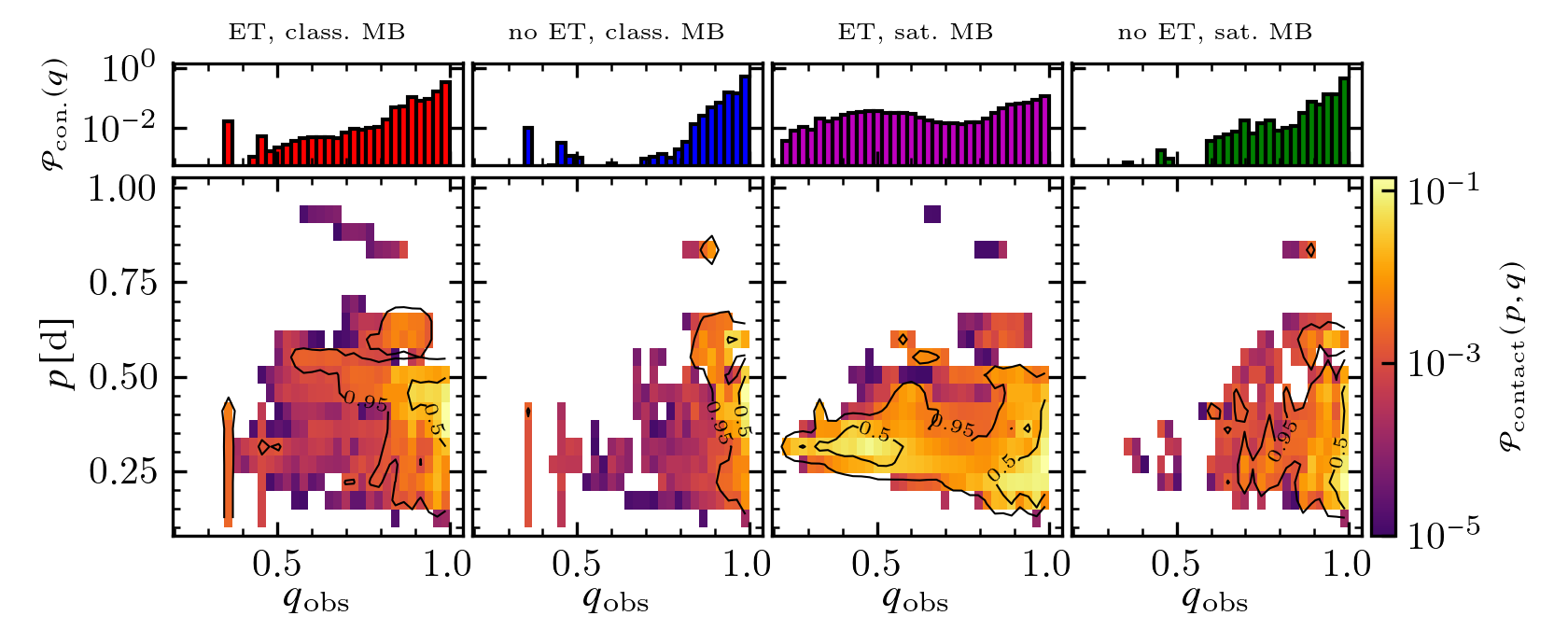}
	\caption{Probability density of our model grids. The top panels show the marginalized mass-ratio distribution, while the lower panels show the two-dimensional ($p, q_{\rm obs}$) distribution. We note that only the models with ET and saturated MB are able to produce systems with mass ratios significantly away from unity.}
	\label{fig:massratios}
\end{figure*}

In Fig.~\ref{fig:massratios}, we show the probability density of contact binaries as a function of the observed mass ratio, $q_{\rm obs} = \min(q, 1/q)$, and the observed period.
We overlay two contours indicating the highest-density intervals of quantiles 0.5 and 0.95.
We note that only the models with ET and saturated MB predict a significant amount of contact systems with $q_{\rm obs} \lesssim 0.5$.
In the others, the majority clusters around a mass ratio of unity.
Furthermore, we expect the peak at $q_{\rm obs} \approx 0.5$ to smear out to lower mass ratios still as the models are evolved through more TROs (akin to the extrapolation of Fig.~\ref{fig:extrapolation}).
Conversely, we emphasize that all models that do not experience TROs are not halted prematurely.
The mass-ratio distributions of the no-ET and/or classical MB grids are the true expected distributions of the whole W UMa population (from first contact to when \Lag{2} overflow occurs).

From Figs.~\ref{fig:taus} and \ref{fig:massratios}, we see that only with the combination of ET and saturated MB are systems produced that live longer than around $10\tau_{\rm KH}$, while keeping the overflow rates low and driving the mass ratio away from unity.
This suggests that MB is significantly weaker than expected from the widely adopted Skumanich law of Eq.~\eqref{eq:mb_rvj}.
\citet{el-badryMagneticBrakingSaturates2022} found similar evidence from studying a sample of detached binaries with periods under $\qty{10}{\day}$.

\subsection{Reduction of magnetic activity by ET}\label{ssec:reduction}
In Sect.~\ref{ssec:ang_mom_tides}, we followed \citet{podsiadlowskiEvolutionarySequencesLow2002} in reducing the MB strength by a factor $\exp(1-0.02/\chi_{\rm conv})$ for tenuous convective envelopes.
To agree with the observations of \citet{ruttenMagneticStructureCool1986} that magnetic activity seems to cease at spectral types earlier than F3 (roughly a $\qty{1.3}{\Msun}$ main-sequence star), \citet{pylyserEvolutionLowmassClose1988} computed that this corresponds to a minimum depth of the convective envelope of $\qty{45e3}{\kilo\meter}$, which will contain roughly 2\% of the stellar mass.

During binary interaction phases, the internal structure adapts to account for any mass and/or energy changes.
In particular, mass donor envelope luminosity decreases as the gas expands when the top layers of the star are removed.
This in turn directly affects the radiative temperature gradient, and thus the criterion where convective instabilities set in.
From the Ledoux stability criterion \citep{ledouxStellarModelsConvection1947}, we have that when the radiative temperature gradient is steeper than the Ledoux gradient, $\nabla_{\rm rad} > \nabla_{\rm L}$, convection will occur.

\begin{figure}
	\centering
	\includegraphics[width=\columnwidth]{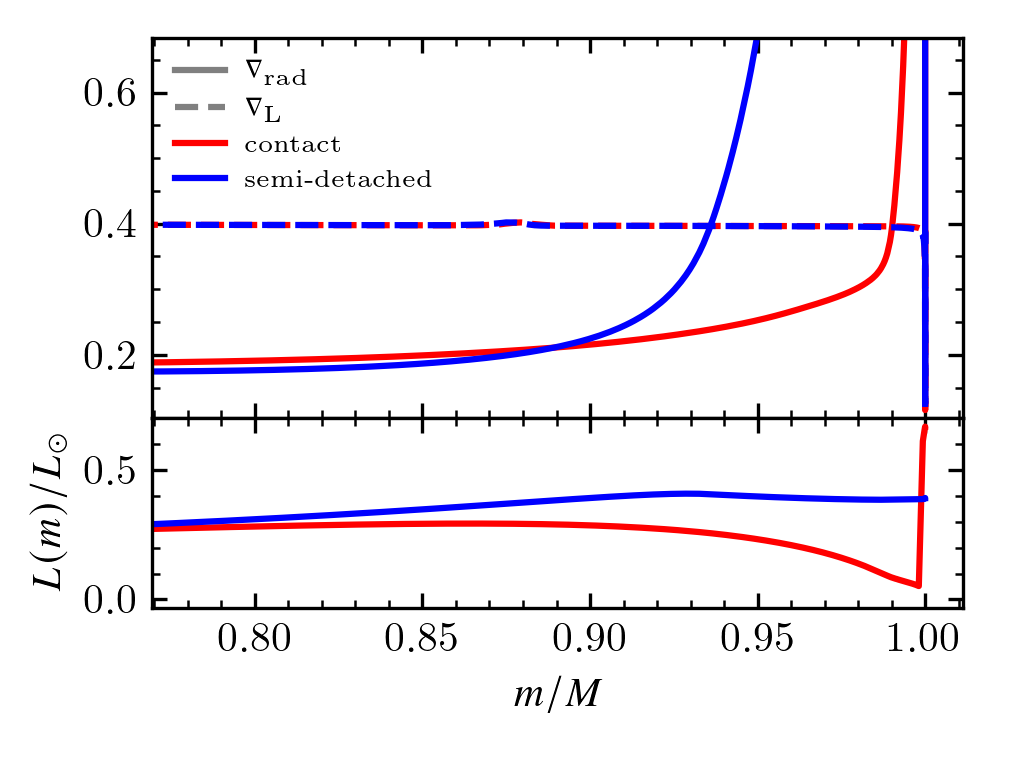}
	\caption{Ledoux ($\nabla_{\rm L}$) and radiative temperature gradients ($\nabla_{\rm rad}$) as a function of the mass coordinate for the lower-mass component in the contact (red) and semi-detached (blue) phases of the TROs. (Bottom:) Luminosity profile of the envelope. During the contact phase, the radiative gradient is suppressed by a decrease in internal luminosity.}
	\label{fig:gradients}
\end{figure}

\begin{figure}
	\centering
	\includegraphics[width=\columnwidth]{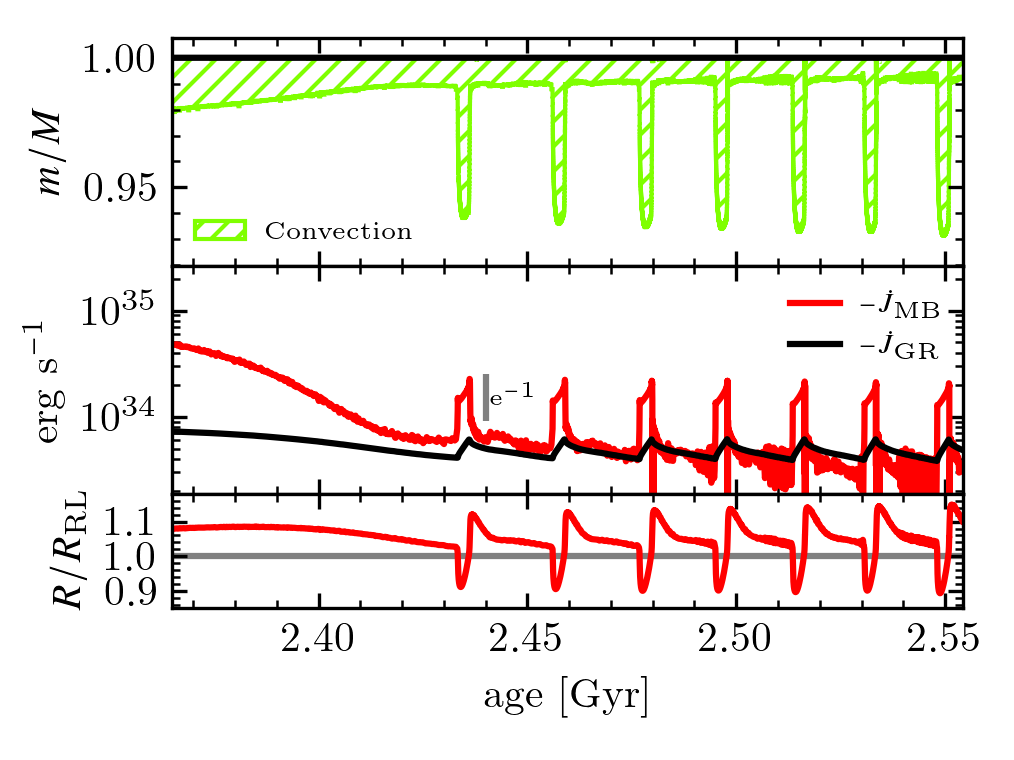}
	\caption{(Top:) Kippenhahn diagram of the primary star of the $M_{1, \rm init} = \qty{1.1}{\Msun}$, $q_{\rm init} = 0.75$, $p_{\rm init} = \qty{0.67}{\day}$ with ET and saturated MB during the first few TROs. (Middle:) Strength of MB and GR AML. (Bottom:) Radius evolution. The drop in MB strength, indicated with the gray vertical bar, is commensurate to the $\exp(-1)$ factor introduced by shallow convective envelopes (see Sect.~\ref{ssec:ang_mom_tides}).}
	\label{fig:kipp}
\end{figure}

We see the changing of convective regions occurring in our models of contact binaries.
In Fig.~\ref{fig:gradients}, we show both $\nabla_{\rm L}$ and $\nabla_{\rm rad}$ for the model with initial parameters $M_{\rm init} = \qty{1.1}{\Msun}$, $q_{\rm init} = 0.75$, $p_{\rm init} = \qty{0.67}{\day}$, showing profiles in the contact (red lines) and semi-detached (blue lines) phases of the TROs.
During the contact phase of the TRO, where the star is the mass donor and energy gainer, its luminosity in the convective layers is severely decreased by binary interaction, and thus its radiative gradient is suppressed, since $\nabla_{\rm rad} \propto L$ \citep[see also][]{stepienLargescaleCirculationsEnergy2009}.
The convective envelope only contains around 1\% of the total mass of the star, while the semi-detached phase, it is around 7\% (see also the top panel of Fig.~\ref{fig:kipp}).

The effect on the strength of MB is shown in Fig.~\ref{fig:kipp}.
The top panel shows the extent of the convective envelope in function of the mass coordinate.
The bottom two panels show that the MB strength is reduced by a factor of around $\exp(1 - 0.02/0.01) = \exp(-1)$, during the contact phases of the TROs.
In this case, MB becomes weaker than AML from GR over the course of several TROs.
This lends extra evidence that contact binaries that engage in binary interaction likely have their MB strength reduced, regardless of whether field saturation due to rotation occurs.
We note that this effect is strong only in stars with masses $M \gtrsim \qty{1.0}{\Msun}$.
At lower mass the convective envelope is deeper and so the fractional reduction in MB strength as a result is expected to be lower.

\subsection{Sensitivity to the ET prescription}
In \citet{fabryModelingContactBinaries2025}, it is shown that for systems evolving on the nuclear timescale, like massive contact binaries during the slow case A MT, the precise prescription of ET is not important.
The envelope always has enough time to thermally adjust before significant evolution occurs.
Any thermally stable model will share this characteristic.
However, the models presented here are clearly not thermally stable, and so it is expected that their behavior changes upon changing the ET prescription.
Here we compare the effects of both extremes in modeling ET.
On the one hand, we model very localized ET at the RL (approximating the discontinuity model of \citealt{shuStructureContactBinaries1976}) and on the other hand ET occurs throughout the all common layers, which models a very baroclinic envelope that reaches a common temperature only at the surface.

\begin{figure}
	\centering
	\includegraphics[width=\columnwidth]{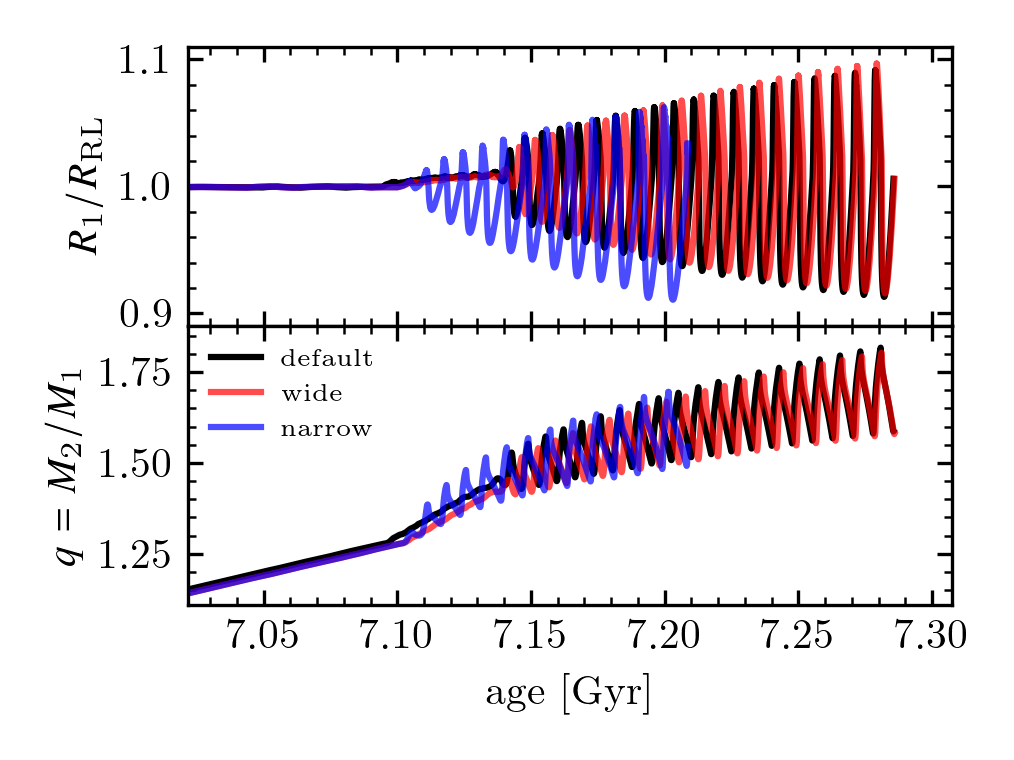}
	\caption{Radius of the primary (Top) and mass-ratio evolution (Bottom) under different prescriptions of the width of the ET region. The narrow model promptly starts TROs when contact is engaged, while wider ET models have a short quasi-stable phase.}
	\label{fig:widths}
\end{figure}

In this section, we consider the model with initial parameters $M_{\rm 1, init} = \qty{1}{\Msun}$, $p_{\rm init} = \qty{1.47}{\day}$, $q_{\rm init} = 0.75$ in the ET grid with saturated MB.
We choose these initial conditions because the default model develops strong TROs with semidetached phases (unlike the example model above).
First, we compute its evolution when ET assumed to occur throughout the whole envelope, without changing its depth below the RL (see Eq.~\ref{eq:et_width}).
Therefore the energy transferring layer will have relative overflows of:
\begin{equation}
	r > -0.02.
\end{equation}
In a second model, we consider a narrow ET region with relative overflows four times smaller than the default of Eq.~\eqref{eq:et_width}:
\begin{equation}
	|r| < 0.005.
\end{equation}
We compare the late-time radius and mass-ratio evolution in Fig.~\ref{fig:widths}.
We find that the secular evolution of this system is relatively unaffected by the width of the ET layers.
All systems develop TROs and evolve away from equal masses, approximately at the same rate.
The largest differences occur before the onset of large TROs.
In the narrow model, TROs develop very quickly after first contact, while the model with a wide ET region is quasi-stable for several thermal timescales before full-scale TROs ensue.
It should also be mentioned that the narrow model terminated prematurely due to convergence issues.
This is associated with the internal luminosity approaching zero at the bottom of the ET region, which is difficult for the solver of \texttt{MESA} to find.
Reinstating the numerical smoothing (see Sect.~\ref{ssec:binary_methods}) might alleviate such issues, but, generally, wider ET regions are more stable from a numerical standpoint.

The extent of the ET process within the envelope thus influences how quickly TROs with semi-detached phases develop.
In Fig.~\ref{fig:example}, we presented a model that does not detach for at least 20 thermal cycles, and the model in Fig.~\ref{fig:kipp} also has an extended contact phases without oscillations before the TROs start.
An in-depth study of the effects of ET prescriptions over the whole population is beyond the scope of the present work, but given the variety of behaviors in the different models presented here, it is warranted.
Furthermore, the amount of ET that is carried internally is uncertain.
\citet{zhouExplanationLightCurveAsymmetries1990} find that surface flows could play an important role in explaining light-curve asymmetries (see also, Fabry \& Pr\v sa, 2025, in prep.), and \citet{stepienLargescaleCirculationsEnergy2009} calculated that such flows would have high heat capacities, such that the internal structure is hardly affected.

\subsection{Constraints on MB strength}
To explain the contact binaries with low mass ratios, our population synthesis imply MB must be considerably weaker than the Skumnanich law.
This is in stark contrast with the results of \citet{vanLowmassXrayBinaries2019}.
They found that in order to explain the high MT rates in some LMXBs with neutron stars, they required a convection-boosted prescription, where the Skumanich law is used, and even boosted to carry a dependence on the convective turnover timescale $(\tau_{\rm conv} / \tau_{\rm conv, \odot})^2$.
One key assumption of the boosted models of \citet{vanLowmassXrayBinaries2019} is that they assume a fully radial field.
This approximation might be appropriate for LMXBs where the accretor is a compact object and thus not distorts the field of the donor.
In the close binaries of \citet{el-badryMagneticBrakingSaturates2022} however, and certainly in the case of contact binaries, the magnetic field is unlikely to be radial because the fields of both components will interact, and more complex morphologies could be created.
This idea is supported by the fact that polars, i.e., CVs with a strongly magnetized white dwarf, have longer orbital periods than non-magnetic CVs \citep{belloniEvidenceReducedMagnetic2020}.
Irrespective of the morphology, observations of chromospheric activity in close binaries suggest that they have enhanced magnetic fields \citep{yuEnhancedMagneticActivity2025}, which can partially cancel out the effect shown in Sect.~\ref{ssec:reduction}.

Recent work by \citet{belloniResolutionParadoxSDSS2025} suggests that in some CVs and AM CVns, MB can, even for very-low-mass donors, be enhanced if the donor is evolved off the main sequence on the sub-giant branch.
This fits into the picture of the (partially) degenerate cores remaining radiative, thus avoiding the fully convective drop off \citep[see also,][]{passConstraintsSpindownFully2022}.
Getting constraints for contact binaries with (slightly) evolved components will be hard, however, as contact binaries are overwhelmingly main sequence-main sequence objects (although systems such as V1309 Sco, with its large period was likely somewhat evolved).
Furthermore, using contact-binary populations as a direct probe for MB strength is challenging because the interplay between MT, ET and nuclear evolution is not yet understood, and angular momentum might not be conserved in these processes.

\section{Conclusions}\label{sec:conc}
In this work, we studied the evolution of low-mass contact binaries under different descriptions of ET and MB.
\par

We computed the expected lifetimes of contact binaries under different descriptions of AML from MB, and showed that the classical MB prescription based on the Skumanich spin-down law would make contact systems merge on the thermal timescale of solar-type stars.
This is line with the findings of \citet{el-badryMagneticBrakingSaturates2022}, who analyzed the period distribution of the slightly larger-period detached binaries.
We presented another argument in favor of reduced MB strength.
Because binary interaction during contact phases has the effect of lowering the radiative temperature gradient in the convective envelope (see also \citealp{stepienLargescaleCirculationsEnergy2009}), contact binaries could have reduced magnetic activity compared to non-interacting counterparts of the same mass.
We noted however that other populations of binaries, such as CVs and AM CVns, experience enhanced MB \citep{vanEvolvingLMXBsCARB2019, belloniResolutionParadoxSDSS2025}, so an investigation into the differences of these populations (and their effects on MB) is needed.
Still, using contact binaries as probes of MB themselves is challenging, given their complex evolution and their propensity to be in triple systems \citep{tokovininTertiaryCompanionsClose2006}.
Instead, we should focus on the detached counterparts that have not interacted yet (and are not in triples, or at most have a very wide third component so that Kozai-Lidov cycles are negligible), such as done in \citealp{el-badryMagneticBrakingSaturates2022}).
\par

Within the context of contact-binary evolution, we showed next that only saturated MB with ET allows systems to evolve toward mass ratios significantly away from unity.
During contact phases, our models develop thermal instabilities as found by \citet{flanneryCyclicThermalInstability1976}, \citet{yakutEvolutionCloseBinary2005} and others.
In the grids with non-saturated MB and/or no ET, the models either merge too quickly or evolve toward equal masses.
\par

The methods developed here allow for a further, comprehensive study of W UMa systems, through the open-source binary-evolution code \texttt{MESA}.
The prescriptions of ET and/or MB can be changed, or the MT efficiency can be adapted to allow for non-conservative MT.
Furthermore, the efficiency and location of internal ET is yet unconstrained.
This might have important ramifications for the fraction of time the binaries spend in the contact vs.~semi-detached phase of TROs, or if TROs happen at all.
Analyzing statistical samples of observed contact-binary light curves could reveal that surface flows play an important role of equalizing effective temperatures of W UMa components (rather than internal currents, see, e.g., \citealp{zhouExplanationLightCurveAsymmetries1990}, \citealp{stepienLargescaleCirculationsEnergy2009}).
This would imply internal ET near the RL equipotential is not as efficient as modeled here, affecting the contact evolution of close binary stars.
\par

\begin{acknowledgments}
	F.M.~and P.~A.~recognize the support from the NSF under grant no.~AST-2306996.
\end{acknowledgments}

\bibliography{../../../bib_exports/massive_stars}
\bibliographystyle{aasjournal.bst}

\end{document}